\documentclass{jps-cp}
\usepackage{txfonts} 

\newcommand{\be}{\begin{equation}}
\newcommand{\ee}{\end{equation}} 
\newcommand{\beq}{\begin{eqnarray}}
\newcommand{\eeq}{\end{eqnarray}}

\newcommand{\p}{\partial}

\newcommand{\bea}{\begin{eqnarray}}
\newcommand{\eea}{\end{eqnarray}}

\def\e{\epsilon}

\def\dcfl{\Delta_{\textsc{cfl}}}

\def \l{\textsc{l}}
\def \r{\textsc{r}}
\def \B{\textsc{b}}
\def \c{\textsc{c}}
\def \f{\textsc{f}}

\def\cfl{{\rm CFL}}
\title{Quark-Hadron Crossover with Vortices}

\author{Chandrasekhar \textsc{Chatterjee}$^{1, a}$, Muneto \textsc{Nitta}$^{1, b}$ and Shigehiro \textsc{Yasui}$^{1, c}$ }

\inst{$^{1}$Department of Physics $\&$ Research and Education Center for Natural Sciences, Keio University,Hiyoshi 4-1-1, Yokohama, Kanagawa 223-8521, Japan}

\email{chandra@phys-h.keio.ac.jp$^a$, nitta(at)phys-h.keio.ac.jp$^b$, yasuis@keio.jp$^c$ }

\recdate{\today}

\abst{The quark-hadron crossover conjecture was proposed as a continuity between hadronic matter and quark matter with no phase transition. It is based on matching of symmetry and excitations in both the phases. It connects hyperon matter and color-flavor locked (CFL) phase of color superconductivity in the limit of light strange quark mass. We study  generalization of this conjecture in the presence of topological vortices. We propose a picture where  hadronic superfluid vortices in hyperon matter could be connected to  non-Abelian vortices (color magnetic flux tubes) in the CFL phase during this crossover.  We propose that three hadronic superfluid vortices must join together to three non-Abelian vortices with different color fluxes with the total color magnetic fluxes canceled out, where the junction is called a colorful boojum.}

\kword{Hadron-quark continuity, CFL phase, Non-Abelian vortices}

\begin{document}
\maketitle

\section{Introduction}

At asymptotically high densities, QCD becomes asymptotically free where the existence  color superconductivity was predicted\cite{Alford:1997zt}. At very high densities strange quark mass can be neglected, the system shows an exact $SU(3)$ flavor symmetry, known as `color-flavor locked phase (CFL)' where underlying symmetry is a diagonal subgroup $SU(3)_{\c +\f}$. The existence of a gap breaks the  $SU(3)_\l \times SU(3)_\r$ chiral symmetry   to the diagonal subgroup $SU(3)_\f$.  It was conjectured from the symmetry structure that there exists no phase transition while  reducing densities (hadronic phase). This is known as `quark-hadron' continuity\cite{Schafer:1998ef}. It is based on the matching of elementary excitations (mainly Nambu-Goldstone modes) and existing global symmetries in both the matter. This conjecture can be  applied to the case of neutron stars, since color superconductivity may exist at the core of neutron stars \cite{Sedrakian:2018ydt} 
and the density decreases  from the core to the crust.  
Neutron stars may realize a continuity between two superfluids; one is in hadronic phase (lower density region) and the other is in the CFL phase \cite{Masuda:2012kf,Baym:2017whm}. Since neutron stars rotate rapidly, 
one should study the continuity of superfluid vortices while going from the CFL to hadronic phase by reducing densities \cite{Alford:2018mqj}.  We reach a conclusion \cite{Chatterjee:2018nxe}
that  smooth connection of vortices from both the phases are possible if there is a  formation of a junction (boojum) \cite{Cipriani:2012hr,Eto:2013hoa} of three hadronic vortices in the hyperon matter with three non-Abelian (NA) vortices in the CFL phase.

\section{Vortices in hadronic  and CFL phases }
In this section we discuss possible vortex configurations in the hadronic and  CFL phases. Here we assume degenerate mass limit for up, down and strange quarks.
In this paper we discuss $\Lambda$ hyperons only, since it  is the lightest one with an 
attractive potential in the nuclear matter. To establish our goal we  consider the most attractive  $\Lambda\Lambda$  pairing (flavor symmetric) in the $^1S_0$ channel \cite{Takatsuka:2000kc, Sedrakian:2018ydt} which  breaks the $U(1)_\B$ baryon number symmetry and  there exist superfluid vortices. 
The vortex ansatz    in the cylindrically symmetric case is given by
$\Delta_{\Lambda\Lambda}(r,\theta) = |\Delta_{\Lambda\Lambda}(r)| e^{i\theta}$,
with the distance $r$ from the center of the vortex and     the (azimuthal) angle $\theta$ around the vortex axis.
 Here $|\Delta_{\Lambda\Lambda}(r)|$  is the  profile function with boundary condition $|\Delta_{\Lambda\Lambda}(0)| = 0, |\Delta_{\Lambda\Lambda}(R)| = \text{gap of the condensate}$, where $R$ is the system
 boundary. The exact nature of profile 
 can be computed from the Ginzburg-Landau (GL) theory in principle. 
 The Onsager-Feynman circulation which is defined as
$
C = \oint \vec v \cdot d \vec l = \frac{2\pi n}{\mu},
$
where $n$ and $\mu$ are the winding number and chemical potential of the condensate, respectively can be computed for
a single $\Lambda\Lambda$ vortex to be $C_{\Lambda\Lambda} = \frac{2\pi }{2\mu_\B}$
where $\mu_\B$ is the chemical potential for a single baryon. Here $\vec v$ is the superfluid velocity at large distance from the core of the vortex and it can be computed from $\nabla\theta$.

The symmetry group of the CFL phase is $U(1)_\B \times SU(3)_{\c} \times SU(3)_{\l}\times SU(3)_{\r}$, where $U(1)_\B$ is the baryon number, $SU(3)_{\l / \r}$ are the left and right chiral symmetries and $SU(3)_\c$ is the color gauge group. The order parameter in the $\cfl$ phase is a matrix 
$\Delta_a{}^i = {\Delta^{\l}}_a{}^i   =-{\Delta^{\r}}_a{}^i$ with a color index $a=1,2,3 \,(r,g,b)$ 
and a flavor index $i=1,2,3 \,(u,d,s)$, 
where
${\Delta^{\l}}_a{}^i  \sim  \e_{abc}\e^{\it ijk} {q^\l}_b^{\it j} \mathcal{C}{q^\l}_c^k, \; {\Delta^\r}_a{}^i  \sim  \e_{abc}\e^{ijk} {q^\r}_b^j \mathcal{C}{q^\r}_c^k$, and $\mathcal{C}$ is the charge conjugation operator. 
The GL formulation of the CFL phase has been derived in  Ref.~\cite{Iida:2000ha}.  The ground state breaks the full symmetry to the diagonal subgroup $SU(3)_{\c+\f}$. In the CFL phase we have two kinds of vortices 1) Abelian vortices 2) NA vortices.
The  Abelian superfluid vortex can be described by the order parameter which can be written as~\cite{Forbes:2001gj,Iida:2000ha} 
$\Delta_{\rm A}(r,\theta) = \dcfl \phi(r)e^{i\theta}\mathbf{1}_3$,
where $\phi(r)$ is a profile function  with boundary conditions $\phi(0)= 0, \phi(R) = 1$, where $R$ is the system boundary. $\dcfl$ is the absolute value of the gap in the CFL phase.  The Onsager-Feynman circulation of Abelian vortices in the CFL phase is given by
$C_{\rm A} = \frac{3\pi}{\mu_\B}$, since  $\mu_{\rm CFL} = \frac{2\mu_\B}{3}$. 
If one proposes the criteria to connect vortices smoothly by matching  the Onsager-Feynman circulation,  a single $\Lambda\Lambda$ vortex cannot connect smoothly to a single Abelian vortex in the CFL phase.
However  three $\Lambda\Lambda$ vortices may join to form one Abelian CFL vortex. 
 
NA vortices are  color magnetic flux tubes and the simplest vortex ansatz 
 is given in Refs.~\cite{Balachandran:2005ev, Nakano:2007dr,Eto:2013hoa} as 
$
\Delta_{ur}(r, \theta)  = 
\dcfl \, {\rm diag}\bigl(f(r)e^{i\theta}, g(r), g(r)\bigr), 
A^{ur}_i(r) =  \frac{1}{3g_s} \frac{\epsilon_{ij} x_j}{r^2} \bigl( 1 - h(r) \bigr) \, {\rm diag}(2, -1,-1), $
with the gauge coupling constant $g_{s}$.
Here the profile functions $f(r)$, $g(r)$ and $h(r)$ can be computed numerically with boundary conditions,
$ f(0) = 0,
\p_r g(r)|_{r=0} = 0, 
h(0) = 1,  f(\infty) = g(\infty) = 1,  
h(\infty) = 0$~\cite{Nakano:2007dr}. 
We define this as an up-red ($ur$) vortex since the $ur$ component of the gap $\Delta$ has a vortex winding. 
The other  two  vortices, down-green ($dg$) and strange-blue ($sb$), are defined by changing the
position of the vortex winding ($e^{i\theta}$) from $\Delta_{11}$ to $\Delta_{22}$ and  $\Delta_{33}$, 
respectively. 
The behavior of the order parameters of these three vortices 
are given as
$\scriptsize \Delta \sim \dcfl e^{i\theta/3} 
 \exp
 \Biggl(
    - i g_s \int_0^\theta \vec{A}\!\cdot\! \mathrm{d}\vec{l}
 \Biggr)
  \mathbf{1}_{3\times 3}, \label{eq:Delta}$
where $A_i$ is the large distance vortex configuration of the gauge field.
 In this case one may derive superfluid velocity from the overall $U(1)_\B$ phase $\theta/3$,
 since  we have to replace ordinary derivative to
covariant derivative in the expression of the current to compute velocity and that cancel the gauge field contribution 
in the current.
The Onsager-Feynman circulation of NA vortices is found to be $C_{\rm N\!A} = \oint v\cdot dl = \frac{\pi}{\mu_\B}$. 
One may notice that it coincides with the circulation of a single $\Lambda\Lambda$ vortex. 
So it can be expected  that a single $\Lambda\Lambda$ vortex would be smoothly connected to a single
NA vortex during the crossover~\cite{Alford:2018mqj}. 
However, in next section we show that this may  be consistent only at the large distance but not at the short distances. 
\section{Existence of boojum}
In this section we discuss  the continuity of phase changes of quark wave functions in the presence of vortices during the crossover. Since the Onsager-Feynman circulation is calculated at large distances, it may loose information about the short distance behavior. Also,  it does not include effect of color-magnetic fluxes, such as the Aharonov-Bohm (AB) phase. Therefore, we  study the matching of the phase changes of quark wave functions after one encirclement of vortices.  In the case of the hadronic phase we may find how quasi-particle of $\Lambda$ changes around a vortex from the Bogoliubov de-Gennes (BdG) equation (see \cite{Chatterjee:2018nxe} for details). It is also understandable from the $U(1)_\B$ baryon number transformation of the $\Lambda$ particle. The wave function acquires a phase around the vortex, given as
$\scriptsize \Psi_{\Lambda}(x)= \left(
\begin{array}{ccc}
 e^{i\frac{\theta}{2}} \psi_\Lambda(r) \\
 e^{-i\frac{\theta}{2}}  \psi_h(r)   \\  
\end{array}
\right) $. 
Here the upper and lower components of $\Psi_{\Lambda}$ are the  wave functions of particle and hole parts of $\Lambda$, respectively. Since these phases are independent of color-flavor indices and $\Lambda = uds$, we conclude that each quasi-quark would acquire a phase as
$\scriptsize |q\rangle_{ai}(x)= \left(
\begin{array}{ccc}
 e^{i\frac{\theta}{6}} q_{ai}(r) \\
 e^{-i\frac{\theta}{6}}  h_{ai}(r)   \\  
\end{array}
\right) $. 
Here  quark's particle and hole wave functions are upper and lower components of $|q\rangle_{ai}$, respectively. So after a full encirclement the quark's phases can be written in a matrix form as
\begin{eqnarray}
 Q_{\Lambda\Lambda}
=
\frac{\pi}{3}
\left(
\begin{array}{ccc}
 +1 & +1 & +1 \\
 +1 & +1 & +1 \\
 +1 & +1 & +1
\end{array}
\right)
 \; {\rm for}\;
q_{a i}
\!=\!
\left(
\begin{array}{ccc}
 u_{r} & d_{r} & s_{r} \\
 u_{g} & d_{g} & s_{g} \\
 u_{b} & d_{b} & s_{b}
\end{array}
\right)
\end{eqnarray}
and the opposite sign for hole components. So these phases of quark wave functions in hadronic phase lives in a $\mathbb{Z}_6$ group. 

 Next let us move to the CFL phase.
First, the phase changes of quasi-quarks around one Abelian CFL vortex is found to be  $\scriptsize Q_{\rm A} = \pi \left(
\begin{array}{ccc}
 +1 & +1 & +1 \\
 +1 & +1 & +1 \\
 +1 & +1 & +1
\end{array}
\right)$.
On the other hand, the phase changes of 
quasi-quarks around one NA vortex are found to be
$\scriptsize q_{ai}(x) = \scriptsize e^{i\frac{\theta}{6}}\left[ P\left(e^{i g \int_0^\theta A\cdot dl}\right)q\right]_{ a i}$.
where the phase is generated by the $U(1)_B$ transformation 
and a holonomy operator in the fundamental representation of the $SU(3)_\c$. 
 The integrals for NA vortices
in an appropriate gauge are found to be
\begin{eqnarray}
 &&q
\rightarrow
e^{i\alpha/6} \,
\mathrm{diag}
\bigl(e^{-i2\theta/3},e^{i\theta/3},e^{i\theta/3}\bigr)
 q =
\mathrm{diag}
\bigl(e^{-i\theta/2},e^{i\theta/2},e^{i\theta/2}\bigr)
 q , \nonumber\\
 &&
 q
\rightarrow
e^{i\theta/6} \,
\mathrm{diag}
\bigl(e^{i\theta/3},e^{-i2\theta/3},e^{i\theta/3}\bigr)
 q =
\mathrm{diag}
\bigl(e^{i\theta/2},e^{-i\theta/2},e^{i\theta/2}\bigr)
 q ,  \nonumber\\
 && q \rightarrow
e^{i\theta/6} \,
\mathrm{diag}
\bigl(e^{i\theta/3},e^{i\theta/3},e^{-i2\theta/3}\bigr)
 q
= \mathrm{diag}
\bigl(e^{i\theta/2},e^{i\theta/2},e^{-i\theta/2}\bigr)
 q 
 \end{eqnarray}
 for the $ur$, $dg$ and $sb$ NA vortices, respectively. 
 After one encirclement, these are 
\begin{eqnarray}
Q_{ur}
=
\pi
\left(
\begin{array}{ccc}
 -1 & -1 & -1 \\
 +1 & +1 & +1 \\
 +1 & +1 & +1
\end{array}
\right),\,
Q_{dg}
=
\pi
\left(
\begin{array}{ccc}
 +1 & +1 & +1 \\
 -1 & -1 & -1 \\
 +1 & +1 & +1
\end{array}
\right),
Q_{sb}
=
\pi
\left(
\begin{array}{ccc}
 +1 & +1 & +1 \\
 +1 & +1 & +1 \\
 -1 & -1 & -1
\end{array}
\right),
\end{eqnarray}
and the opposite sign for the hole components.
In this case, the phases constitute a $\mathbb{Z}_2$ group and  they are gauge invariant. Since  $e^{\pm i \pi} = -1$, all three NA vortices result in the overall minus sign to the wave functions. We call these phases as `generalized AB phases'. So after these computations we find that the phase changes of quarks for  one $\Lambda\Lambda$ and one NA vortex after a full encirclement do not match,   or in other words 
$e^{i(Q_{\Lambda\Lambda})_{ai}} \ne e^{i (Q_{\rm NA})_{ai}}$, with $Q_{\rm NA}=\{Q_{ur}, Q_{dg}, Q_{sb}\}$. 
This mismatch can be understood  from the difference in the group structure (the $\mathbb{Z}_6$ group in the hadronic side and the $\mathbb{Z}_2$ group in the CFL side) involved in these processes.   Now let us try to understand the actual structure. In the hadronic side, let us take a bundle of three $\Lambda\Lambda$ vortices and  compute the phase changes of quarks when encircling them together. 
 On the other hand  in the CFL side  if we consider generalized AB phase of quarks around the bundle of any of three  NA vortices we find $e^{i3{\left(Q_{\Lambda\Lambda}\right)}_{ai}} =  e^{i 3 {\left(Q_{\rm NA}\right)}_{ai}}=  -1$. Since  the phase changes  around a single $\Lambda\Lambda$ vortex cannot match to the phase changes around a single NA vortex during the crossover period, a bundle of three $\Lambda\Lambda$ vortices must join themselves to a bundle of three NA vortices.  Only possibility that satisfies this condition is that the joining region is composed of one Abelian CFL vortex, since the phase changes of quarks around a single Abelian vortex are $(Q_{\rm A})_{ai} = \pm\pi$ irrespective to color and flavor. However, we know that a single Abelian CFL vortex can decay to three different ($ur, gd, sb$) NA vortices because of minimization of energy \cite{Nakano:2007dr}. So we conclude that only possibility which is left for a boojum construction is following: three  $\Lambda\Lambda$ vortices join together to one Abelian CFL vortex during crossover which is further splitting into  three different ($ur, gd, sb$) NA vortices. In other words we may write it in terms of one equation as
\begin{eqnarray}
3 Q_{\Lambda\Lambda} = Q_{\rm A} = Q_{ur} + Q_{dg} + Q_{sb}.
\end{eqnarray}
 
 \section{Conclusion}
 We have  discussed the  hadron-quark continuity in the presence of vortices based on Ref.~\cite{Chatterjee:2018nxe}. At first we have introduced different vortices in hadronic phase and CFL phase. Then we have defined a condition for continuation of vortices from the hadronic phase to the CFL phase, implying the matching of phases of quasi-quark wave function in the presence of vortices. We find that this condition would be satisfied if there exists a junction of vortices during the crossover.
Future directions will include NA statistics of NA vortices \cite{Yasui:2010yh} 
due to fermion zero modes \cite{Yasui:2010yw},
inclusion of electromagnetic interaction \cite{Vinci:2012mc},
the monopole confinement \cite{Eto:2011mk}, 
and a possible connection to a topological order \cite{Cherman:2018jir}.

 This work is supported by the Ministry of Education, Culture, Sports, Science (MEXT)-Supported Program for the Strategic Research Foundation at Private Universities ``Topological Science" (Grant No. S1511006). C. C. acknowledges support as an International Research Fellow of the Japan Society for the Promotion of Science (JSPS) (Grant No: 16F16322). This work is also supported in part by JSPS Grant-in-Aid for Scientific Research (KAKENHI Grant No. 16H03984 (M. N.), No. 18H01217 (M. N.), No. 17K05435 (S. Y.)), and also by MEXT KAKENHI Grant-in-Aid for Scientific Research on Innovative Areas ``Topological Materials Science" No. 15H05855 (M. N.).


%


\end{document}